**Mikhail Fomichev, Max Maass and Matthias Hollick**
*Secure Mobile Networking Lab, Technische Universität Darmstadt, Darmstadt, Germany*

**Editors: Aruna Balasubramanian and Lin Zhong**

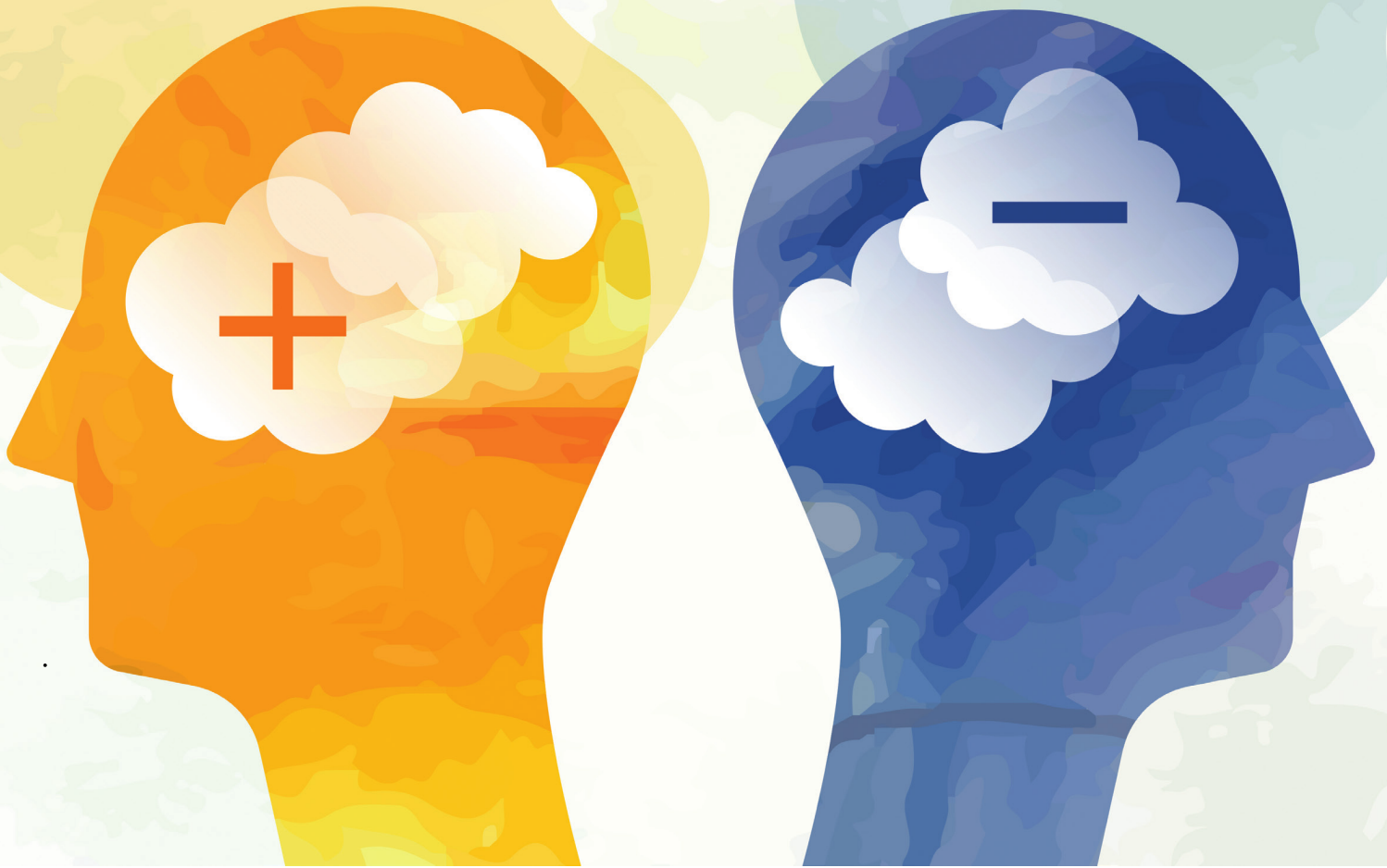

# ZERO-INTERACTION SECURITY—TOWARDS SOUND EXPERIMENTAL VALIDATION

Illustration, istockphoto.com

Reproducibility and realistic datasets are crucial for advancing research. Unfortunately, they are often neglected as valid scientific contributions in many young disciplines, with computer science being no exception. In this article, we show the challenges encountered when reproducing the work of others, collecting realistic data in the wild, and ensuring that our own work is reproducible in turn. The presented findings are based on our study investigating the limits of zero-interaction security (ZIS)—a novel concept, leveraging sensor data collected by Internet of Things (IoT) devices to pair or authenticate devices. In particular, we share our experiences in reproducing five state-of-the-art ZIS schemes, collecting a comprehensive dataset of sensor data from the real world, evaluating these schemes on the collected data, and releasing the data, code, and documentation to facilitate reproducibility of our results.

In our discussion, we outline general considerations when conducting similar studies and give specific examples of technical and methodological issues that we experienced. We hope that our findings will raise awareness about the importance of reproducibility and realistic datasets in computer science and inform future research.

Zero-interaction security (ZIS) allows pairing or authenticating Internet of Things (IoT) devices without user involvement using sensor readings of their ambient environment. The rationale behind ZIS is that *colocated* devices, residing in the same physical space such as a room will observe similar events happening in their environment via on-board sensors (e.g., door opening, people speaking, etc.). Using similarity of their sensor readings, devices can either establish a shared secret key (pairing) or one device can prove its proximity to another device (authentication). The main advantages of ZIS are high usability as no human interaction is required, as well as scalability and deployability as most IoT devices are equipped with sensors. However, existing ZIS schemes were evaluated on different and sometimes unrealistic datasets (e.g., collected with few devices in a controlled environment), making them impossible to compare and calling their real-world utility and security into question.

In our study [1], we reproduced and compared five ZIS schemes, requiring us to collect seven different sensor modalities: audio, temperature, humidity, pressure, luminosity, Wi-Fi, and Bluetooth beacons. To evaluate the security of the schemes under realistic conditions, we collected sensor data from three real-world scenarios (car, office and mobile) using a representative number of heterogeneous sensing devices (up to 25) distributed to reflect potential IoT functionality (e.g., on a display, under the ceiling, etc.), running from several hours to multiple days, which is typical in the IoT. Our results showed that four out of the five schemes experienced significantly degraded security on our datasets [2].

Reproducing the work of others, collecting data and releasing it with the codebase were the main contributions of our study [1], which are often considered "not challenging" or "not novel" enough by reviewers, and hence, fall largely into the category of (almost) unpublishable results per se. In this article, we want to illustrate the difficulties we faced in our study by first describing the challenges encountered when *reproducing* five ZIS schemes such as the absence of the source code and documentation, ambiguities, and unspecified parameters. Second, we elaborate on challenges faced in the *data collection* starting from building a realistic setup to issues hindering reliable data collection such as power, connectivity, and fault tolerance. Third, we describe our experience in *processing collected data* such as ensuring its quality (e.g., identifying erroneous data) and working with large datasets. Finally, we outline best practices we followed for data release.

### REPRODUCING PUBLISHED ALGORITHMS

Researchers commonly differentiate between *reproducibility* (being able to rerun the same code on the same data and obtain the same results) and *replicability* (being able to write a new implementation of the proposed algorithm from scratch, running it on the same data, and obtaining the same results) [3]. In our study, we found both to be impossible, as none of the papers had published their source code or data, which is common in computer science research [4]. After requesting access to code and data via e-mail, one team of authors provided us with their code (but no data), and another team provided us with their data. In the latter case, we were still unable to reproduce their results using the machine learning tool *Weka* that the authors had employed, likely due to different default parameters of machine learning algorithms in different versions of *Weka*, a problem anticipated by Benureau and Rougier [3]. The other authors did not respond to our requests or denied us access to the code and data due to intellectual property and privacy concerns.

This left us with the task of re-implementing all five schemes from scratch, based on the information given in the publication. This effort was hampered by ambiguous descriptions of parts of the algorithms, underspecified behavior for edge cases and, in some cases, missing values for system parameters (e.g., threshold values, sampling rates). We resolved these issues and validated our interpretations of the algorithms in communication with the original authors. However, due to the lack of original datasets, we were unable to replicate the results from the original papers.

To allow for a fair comparison between the five schemes, we decided to collect our own dataset, which will be described in the next section.

## DATA COLLECTION, PROCESSING AND RELEASE

In this section, we first describe our experiences in building a reliable data collection platform, highlighting major issues we faced when deploying it in realistic environments. We then elaborate on challenges when processing our large dataset and, finally, we summarize the main points to consider when releasing data and code. It goes without saying that for large data collection studies involving human subjects (e.g., audio data collection) an institutional review board (IRB) approval needs to be sought, which we recommend doing well in advance as the process may take several months in complex cases.

### Data Collection

In our study, [1] we collected data from three scenarios: connected car, smart office, and mobile. In each scenario, we deployed multiple sensing devices to represent realistic IoT environments—each device was placed in a spot reflecting potential IoT functionality such as under the ceiling (e.g., smart light) or inside a trunk (e.g., smart sensor). Each scenario differs in terms of types of sensing devices used, their mobility and duration of data collection, posing different challenges. In the car scenario, we equipped two cars with six homogenous static devices each, collecting data during a four-hour trip. In the office scenario, we equipped three offices with eight homogenous static devices each, and collected data for one week. In the mobile scenario, we used heterogeneous sensing devices, both statically deployed and carried by users, and collected data for eight hours. Overall, we used four types of sensing devices: a Raspberry Pi 3 with attached TI SensorTag and a Samson Go USB microphone (Pi+Tag+Mic), a Samsung Galaxy S6 smartphone, a Samsung Gear S3 smartwatch, and a RuuviTag. Figure 1 shows examples of sensing devices deployed in our scenarios. In the following, we elaborate on power, connectivity, fault tolerance and testing issues experienced during data collection, and solutions to these issues found.

### Power

While smartphones, watches and RuuviTags have built-in batteries, customized sensing devices such as Pi+Tag+Mic, need an external power supply, accommodating the power consumption of both the attached peripherals and processes running on the main board. We empirically found that a power supply rated below 2.1A current (at 5V) leads to unpredictable behavior of Pi+Tag+Mic devices, hindering reliable data collection and emphasizing the need to carefully choose the power supply for customized sensing devices.

In the office scenario, we used Pi+Tag+Mic devices, which were required to run one week non-stop, making the use of mains supply an obvious choice. However, we could not use AC power adapters due to their insufficient cable length as many devices were placed in spots (e.g., under the ceiling) without power sockets in proximity. Thus, we considered two other alternatives to deliver power over 5-10 meter distances: USB cables and Power over Ethernet (PoE) supplied via external adapters. While the former is much easier to deploy, we found that USB cables experience substantial voltage drops, resulting in insufficient power delivered to Pi+Tag+Mic devices, again hindering reliable data collection. PoE, on the other hand, did not suffer from voltage drops and additionally brought connectivity to the devices (see "Connectivity" section), making us favor this option.

In the car scenario, we also used Pi+Tag+Mic devices; however, the shorter duration of data collection, absence of mains supply and difficulties using bulky cables to deploy sensing devices inside a car motivated the use of portable power banks (10000 mAh). We used the same power banks in the mobile scenario to power a user-carried Raspberry Pi 3, capturing data from RuuviTags, showing that portable power supplies are suitable for several-hour data collections conducted with mobile devices, or when delivering power to static devices over wires is impractical.

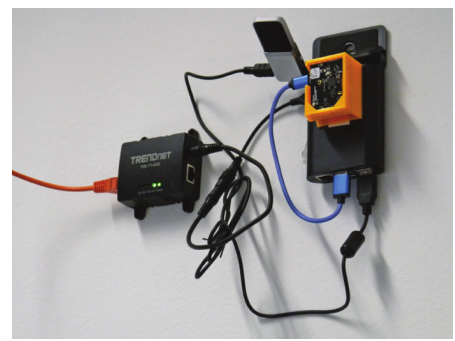

(a)

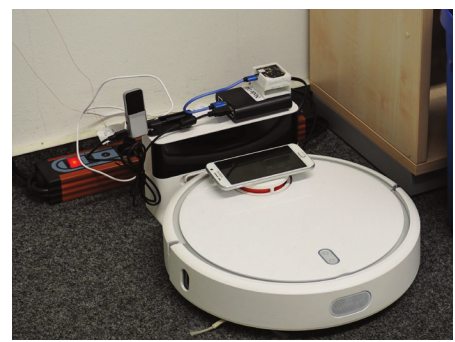

(b)

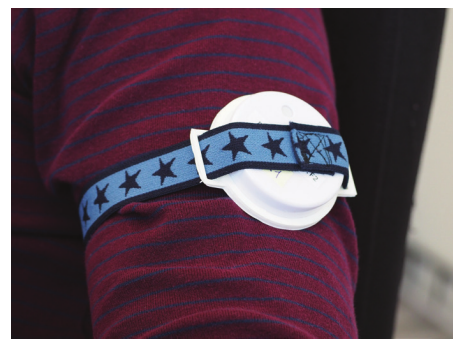

(c)

**FIGURE 1.** Sensing devices deployed:
(a) Pi+Tag+Mic on a wall powered over PoE
(b) Samsung S6 on top of a robot vacuum cleaner (c) RuuviTag on the upper arm.

### Connectivity

In data collections, devices often need different types of connectivity (e.g., to a core network or between each other) for

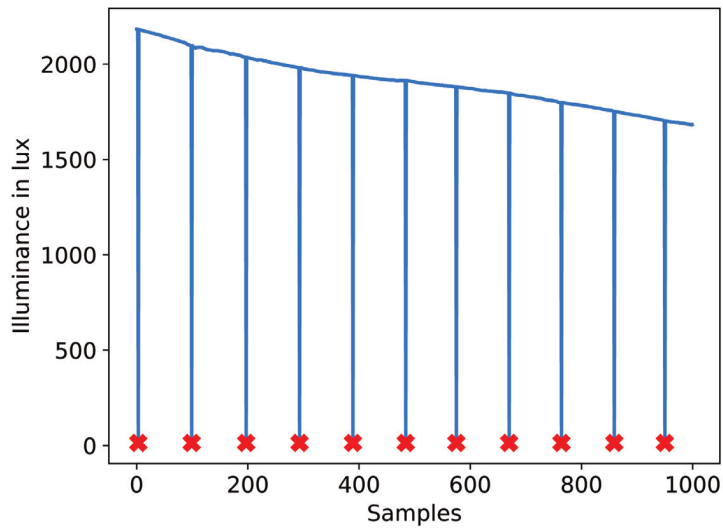

**FIGURE 2.** Example of faulty illuminance values being periodically delivered by SensorTag

> WE HOPE OUR EXPERIENCE CAN HELP RAISE AWARENESS ABOUT THE CHALLENGES RESEARCHERS FACE WHEN ATTEMPTING TO REPRODUCE THE WORKS OF OTHERS AND COLLECTING DATA IN THE WILD

purposes such as time synchronization, monitoring, and data delivery. In our scenarios, all sensing devices needed access to the Internet to perform the NTP update, facilitating the synchronous start of data collection and correct timestamping of sensor readings. We found that zone-specific NTP servers rate limit requests from the same network, so if the number of devices is above 20, we recommend either setting up a dedicated NTP server or spreading NTP requests over time.

Compared to a one-time NTP update, data delivery and data collection monitoring require permanent connectivity between the devices. In realistic environments, connectivity is affected by interferences caused by sensing devices themselves and neighboring devices communicating in the same frequency band. In our scenarios, we observed connectivity drops between Wi-Fi and Bluetooth devices communicating in the 2.4 GHz band, which we attribute to the overloaded spectrum. This caused occasional drops of the Bluetooth link between a SensorTag and Raspberry Pi, permanently terminating sensor data delivery from the SensorTag. We observed similar connectivity drops between sensing devices and a Wi-Fi access point (AP), hindering the use of Wi-Fi to remotely access devices for monitoring. These examples show that wireless connections should not be assumed reliable in realistic environments, and wires should be used instead if reliable connectivity is critical.

**Fault Tolerance**
Fault tolerance is indispensable to ensure reliable data collection. In realistic environments with distributed sensing devices, it is important to monitor liveness of data collection. The easiest way is to remotely access the devices, however, this can either be infeasible (see "Connectivity" section) or undesirable (e.g., security/privacy concerns), making visual inspection a viable alternative. In smartphones and watches, visual inspection is easy to implement due to available user interfaces, however, customized sensing devices such as Pi+Tag+Mic often lack user interfaces, making the use of LEDs imperative to visually monitor liveness of data collection. We leveraged this observation by shutting down Pi+Tag+Mic devices (LEDs go off—easy to notice) in cases of critical data collection errors.

Liveness detection can be coupled with recovery procedures, increasing reliability of data collection. For example, a connectivity drop between a SensorTag and Raspberry Pi (see "Connectivity" section) terminates the process, fetching data from the SensorTag. We thus introduced a watchdog process, continuously monitoring the data-fetching process and restarting it if the process was terminated.

In a distributed setup with many devices, it makes sense to implement a scheduled start-up of data collection. If the data collection is interrupted (e.g., Pi+Tag+Mic powers off), a manual intervention becomes unavoidable. To facilitate a seamless restart of data collection, two points need to be considered: first, the restart must be fast, requiring minimum user interaction such as unplugging a device and plugging it back in or relaunching the data collection app, second, the data collected before the interrupt must be saved separately and not be overwritten by newly collected data.

**Testing**
Testing a data collection platform carefully allows identifying many error cases and ensuring reliability during the real experiment. Here, we outline three crucial points for testing derived from our experience, providing concrete examples of problems encountered and pitfalls to avoid.

First, a data collection platform must be tested for realistic deployment time, corresponding to the duration of actual data collection. This allows identifying, memory leaks, durability of power supplies, and file size limitations—we empirically found a 4 GB WAV size limit imposed by the standard, making us adopt the FLAC format instead. Second, a data collection platform must be thoroughly tested in the exact environment it will be deployed in. For example, we undertested our platform in running cars before collecting data in them, resulting in incorrectly chosen microphone settings, ruining audio recordings as they were saturated by the engine hum, and necessitating a repeat of the experiment.

Third, a data collection platform must be tested with a realistic number of devices, running under realistic loads. Using this principle, we identified a number of problems related to interference and overloaded 2.4 GHz spectrum. For example, before opting for Samson Go microphones, we tried several more affordable alternatives, all of which suffered from interference caused by communicating SensorTags, making the quality of audio recordings unacceptable. In the case of the overloaded spectrum, we found that Wi-Fi captures (i.e., scanning visible APs) crashed on Raspberry Pi and Samsung S6 devices, freezing the whole Wi-Fi interface, indicating that there might be a serious flaw in the Wi-Fi stack of Linux-based devices.

**Data Processing**

Regarding data processing, we elaborate on two main points: first, ensuring the quality of collected data, and second, dealing with large datasets.

Having collected the data, one must ensure its quality, which can be affected by devices stopping recording (and manually restarted), faulty sensor readings, and sampling drift. If the restart of data collection is properly implemented, the sensor data only needs to be stitched, which is straightforward for most modalities except audio. If an audio recording is not terminated correctly, the resulting audio file may become corrupted due to missing file headers. We experienced such cases and used binary hex editors to manually craft audio file headers, completely restoring the audio recording; for stitching audio recordings we used *Audacity*.

Sometimes sensors deliver erroneous readings, which need to be identified and excluded. To do so, we plotted the collected sensor data and visually inspected it. This turned out to be a very powerful tool to spot outliers (see Figure 2) and missing data. This type of sanity checks suffices for most of the modalities except audio. In audio recordings made by heterogeneous devices, we observed non-negligible sampling drift (see Figure 3) caused by internal clock offsets of different devices, despite the synchronous start of audio recordings. To remedy this, we found the lag between heterogeneous audio recordings and applied the time-stretching effect in *Audacity*.

Similar to prior research [6], we found *sensor bias* among heterogeneous devices,

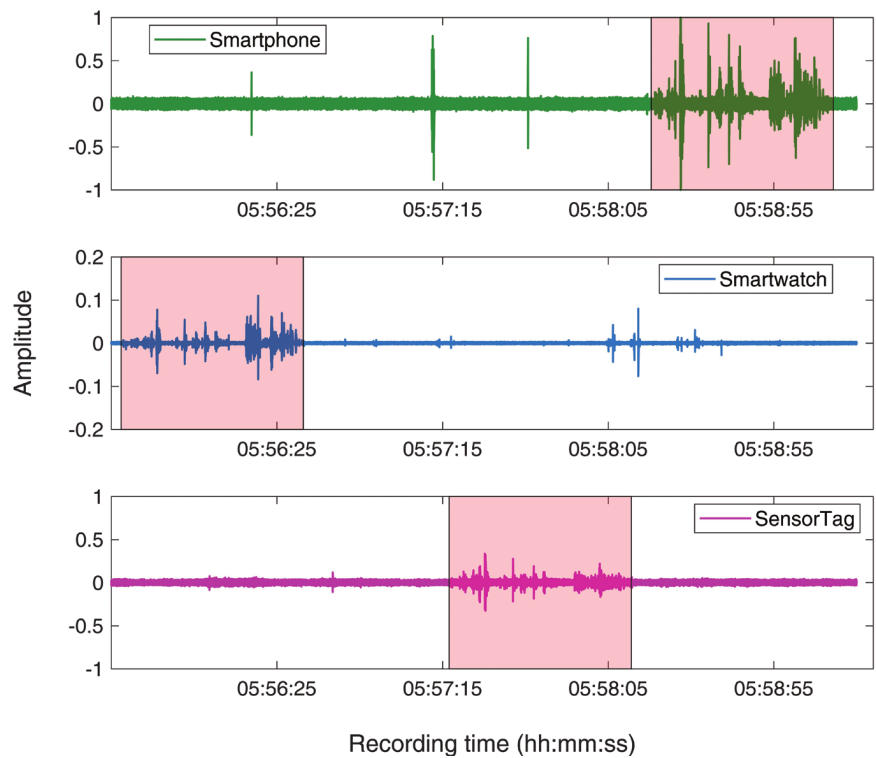

**FIGURE 3.** Sampling drift between heterogeneous devices: insignificant at the beginning of audio recording, it becomes prominent over the course of several hours (highlighted regions indicate the same part of audio recording)

most notably in barometric pressure readings. We also observed *sampling rate instability* among smartphones of the same model, especially for IMU sensors (tested with Galaxy S6, Nexus 5X, and 6P), with the actual sampling rate deviating by up to 10% from the set value. The sampling rate instability can hinder data collection, for example, light sensor readings on Nexus 5X and 6P missed the advertised sampling rate (5 Hz) by a wide margin, going as low as 0.2 Hz and, thus, preventing us from using these smartphones in our study. The sampling rate instability was less prominent on smartwatches and insignificant on SensorTags. This demonstrates that although modern smartphones contain powerful sensors, they can be unsuitable for scientific measurements [7].

After sanity checking the collected data, we ended up with the dataset of 239 GB to compute on. To deal with such large data in a reasonable time, we followed known best practices, which are often overlooked.

First, we ensured sufficient resources with access to our institution's high-performance cluster. To leverage full potential of the cluster, the code running on it needs to be customized to work in a highly parallel environment. When running highly parallel computations, one must consider licensing issues if commercial software is used. For example, to work around shortage of Parallel Computing Toolbox licenses in MATLAB, we used the MATLAB compiler, allowing us to create a standalone application, which can be launched royalty-free on an arbitrary number of machines.

Second, we tried avoiding redundant operations on large datasets. This can be done by using stateful systems such as *Jupyter Notebook*, loading data once and keeping it in RAM afterwards. In the case of time-consuming computations, we recommend liberally caching intermediate results to files, which can significantly speed up reruns of the evaluation.

Third, we applied compression to reduce storage and computing requirements.

Sensor data is often redundant (e.g., slowly changing modalities such as temperature), showing great potential for compression. We leveraged this observation by counting and deduplicating identical instances and assigning them weights in the dataset used to train the machine learning classifiers, reducing its size from 81 GB to 600 MB and decreasing training time by orders of magnitude.

**Data and Code Release**

When releasing a dataset, several issues need to be considered. First and most importantly, the data needs to be sufficiently anonymized to protect the study subjects and comply with legal requirements. If the dataset will be released, this should be communicated clearly to the study participants as part of the informed consent process. We decided to keep the audio recordings from two of our three scenarios private and only make the audio from the third scenario available to others upon request. We took precautions to limit the privacy impact of this release before and during the recording, and obtained consent from all involved parties, including our IRB. Wi-Fi and Bluetooth device identifiers were replaced with pseudonyms. The other parts of the dataset were deemed to be non-critical, as they did not contain any sensitive information.

Once the data has been cleaned and anonymized, an appropriate data repository needs to be chosen. Prior work discussed this issue in more detail [5]. As no fitting specialized repository for zero-interaction security data exists, we chose the noncommercial general-purpose repository *Zenodo*. To ensure that data can be selectively downloaded, we split our dataset into several parts (raw data, processed data, and results for each of the scenarios). Where our data exceeded the size limits of Zenodo, we hosted it on Google Drive and created a *stub* dataset on Zenodo, which contained a link to the Google Drive folder and a list of file hashes to verify the download. We also created an index dataset that contains links to all individual datasets [2].

To release the code, we created a public GitHub repository, chose an Open Source license, and linked it to Zenodo to obtain a DOI for it. This allows others to use our reference implementations of the algorithms under test in their own research. Using a version control system like Git also allowed us to annotate our result files with the exact version of the code that generated them, as defined by the Git commit identifier, and the hashes of the input files. Together with a list of the exact versions of all libraries, this makes it possible for others to exactly reproduce our results, as recommended by Benureau and Rougier [3].

## CONCLUSION

In this paper, we discussed the challenges we encountered when reproducing and validating five state-of-the-art Zero-Interaction Security (ZIS) schemes on a realistic dataset [1]. They included problems in understanding and replicating the published schemes due to ambiguous descriptions and a lack of published source code and datasets, issues with the data collection and processing, and the subsequent release of the dataset. We also discussed a selection of best practices to help others overcome these challenges. We hope our experience can help raise awareness about the challenges researchers face when attempting to reproduce the works of others and collecting data in the wild. We encourage researchers to release source code and data [3] to make this process easier, allowing others to advance the state of knowledge by attempting to confirm or invalidate their results. ■

**Acknowledgement**

This work has been co-funded by the DFG within CRC 1119 CROSSING and CRC 1053 MAKI projects, and by the RTG 2050 "Privacy and Trust for Mobile Users."

**Mikhail Fomichev** is a PhD candidate at the Secure Mobile Networking Lab in the Computer Science Department of Technische Universität Darmstadt, Darmstadt, Germany. His main research interests are practical context-based pairing and authentication solutions, enhancing security and privacy in the Internet of Things.

**Max Maass** is a PhD candidate at the Secure Mobile Networking Lab in the Computer Science Department of Technische Universität Darmstadt, Darmstadt, Germany. His research interests include privacy and security for the Internet of Things, and the use of transparency-enhancing technologies to make invisible tracking systems visible to the end user.

**Matthias Hollick** heads the Secure Mobile Networking Lab in the Computer Science Department of Technische Universität Darmstadt, Darmstadt, Germany. After receiving his PhD degree there in 2004, he has been researching and teaching at TU Darmstadt, Universidad Carlos III de Madrid, and the University of Illinois at Urbana Champaign. His research focus is on resilient, secure, privacy-preserving, and quality-of-service-aware communication for mobile and wireless systems and networks.